\documentclass[12pt,epsf,qsymbols]{article}
\pdfoutput=1
\usepackage[utf8]{inputenc}
\usepackage[normalem]{ulem}
\usepackage[english]{babel}
\usepackage{tabularx}
\usepackage{array}
\usepackage{soul}
\usepackage{graphics}
\usepackage[pdftex]{graphicx}
\usepackage{psfrag}
\usepackage{epsfig}
\usepackage{subfig}
\usepackage{mathtools}
\usepackage{amssymb}
\usepackage{bbold}
\usepackage{setspace}
\usepackage{rotating}
\usepackage{colortbl}
\usepackage{longtable}
\usepackage{slashed}
\usepackage{braket}
\usepackage{lineno}
\makeatletter
\usepackage{textcomp}
\usepackage[usenames,dvipsnames]{xcolor}
\usepackage{relsize}
\usepackage{cite}

\usepackage{float}

\usepackage{bbm}
\usepackage{bm}
\usepackage{enumerate}
\usepackage{amsmath}
\usepackage{verbatim}

\usepackage{hyperref}
\hypersetup{colorlinks,citecolor=nicegreen,linkcolor=niceblue}
\hypersetup{colorlinks=true}

\usepackage{pifont}

\usepackage{makecell}

\usepackage{arydshln}

\allowdisplaybreaks

\setlongtables

\newcommand{\bea}{\begin{eqnarray}}
\newcommand{\eea}{\end{eqnarray}}
\newcommand{\beq}{\begin{equation}}
{
\newcommand{\eeq}{\end{equation}}
\newcommand{\ec}{\end{center}}
\newcommand{\bc}{\begin{center}}

\newcommand{\pdir}{p\kern -5.2pt\raise 0.2ex\hbox {/}}

\newcommand{\vdir}{v\kern -5.75pt\raise 0.15ex\hbox {/}}
\newcommand{\kdir}{k\kern -5.75pt\raise 0.15ex\hbox {/}}
\newcommand{\epsdir}{\epsilon\kern -5.0pt\raise 0.15ex\hbox {/}}
\newcommand{\bvdir}{\bar{v}\kern -5.75pt\raise 0.15ex\hbox {/}}
\newcommand{\Ddir}{D\kern -7.75pt\raise 0.20ex\hbox {/}}
\newcommand{\Adir}{A\kern -7.75pt\raise 0.20ex\hbox {/}}
\newcommand{\ldir}{l\kern -5.0pt\raise 0.2ex\hbox{/}}
\newcommand{\varepsdir}{\varepsilon\kern -5.5pt\raise 0.15ex\hbox{/}}

\makeatother

\definecolor{niceblue}{rgb}{0.15,0.15,0.6}
\definecolor{nicegreen}{rgb}{0.1,0.5,0.1}
\definecolor{Red}{rgb}{1.,0.,0.}

\definecolor{Green}{rgb}{0.2,.7,0.2}

\begin{document}
\unitlength = 1mm

\thispagestyle{empty} 
\begin{flushright}
\begin{tabular}{l}
\end{tabular}
\end{flushright}
\begin{center}
\vskip 3.4cm\par
{\par\centering \textbf{\LARGE  
\Large \bf Seeking leptoquarks in the $\bf t\bar{t}$ plus missing energy \\[0.2em] channel at the high-luminosity LHC}}
\vskip 1.2cm\par
{\scalebox{.85}{\par\centering \large  
\sc Natascia Vignaroli}
{\par\centering \vskip 0.7 cm\par}
{\sl 
Istituto Nazionale di Fisica Nucleare, Sezione di Padova, \\ via Marzolo 8, I-35131 Padova, Italy}\\
{\par\centering \vskip 0.2 cm\par}

{\vskip 1.65cm\par}}
\end{center}

\vskip 0.85cm

\begin{abstract}
 The $t\bar t$ plus missing energy channel is one of the most efficient to detect third-generation leptoquarks (LQs). It offers an important test to models which explain flavor anomalies in $B$ meson decays. We outline a search strategy in the channel, relying on tagging the tops and on observables constructed out of the tops, and we assess the reach on LQs of the future high-luminosity LHC program. We find that with 3 ab$^{-1}$ a vector (scalar) LQ decaying 50\% (100\%) to top and neutrino can be excluded up to masses of 1.96 TeV (1.54 TeV). We also indicate several observables that, in case of a future discovery in the channel, can be used to distinguish a scalar LQ from a vector LQ. The implications of our findings to models addressing the recent flavor anomalies are finally discussed.
\end{abstract}

\vskip 4cm
\thispagestyle{empty} 
\begin{flushleft}
\begin{tabular}{c}
\href{mailto:vignaroli@pd.infn.it}{\small{vignaroli@pd.infn.it}}
\end{tabular}
\end{flushleft}

\newpage
\setcounter{page}{1}
\setcounter{footnote}{0}
\setcounter{equation}{0}
\noindent

\renewcommand{\thefootnote}{\arabic{footnote}}

\setcounter{footnote}{0}


\newpage

\section{Introduction}
\label{sec:intro}

Leptoquarks (LQs), which are hypothetical particles carrying both lepton and baryon number, appear in a variety of theories beyond the Standard Model (BSM), as Pati-Salam model \cite{Pati:1974yy}, grand unification theories \cite{Georgi:1974sy} and BSM composite dynamics \cite{Pelaggi:2017wzr}.   
Recently, third generation LQs caught a special attention from the high energy physics community, since they represent the best candidates \cite{Fajfer:2012jt, Alonso:2015sja, Fajfer:2015ycq, Barbieri:2015yvd, Becirevic:2016yqi, Cai:2017wry}  to explain anomalies in flavor physics observed by  experiments on $B$-meson decays:  Belle \cite{Huschle:2015rga, Hirose:2016wfn, Sato:2016svk, Abdesselam:2016cgx}, BaBar \cite{Lees:2012xj, Lees:2013uzd} and by LHCb \cite{Aaij:2015yra, Aaij:2014ora, 1705.05802}. In particular, the experiments find the indication  of  lepton flavor universality violation in the rather clean ratio observables $R_{D^{(*)}}$, at about 4$\sigma$ level (by combining the results of the different experiments), and $R_{K^{(*)}}$, again at about 4$\sigma$ level. 
 It is really appealing that the anomalies can be explained simultaneously by models with LQs in the TeV range \cite{DiLuzio:2017chi}, thus in the reach of the LHC. Furthermore, some models with LQs can also address the discrepancy from the SM in the muon magnetic moment \cite{Bauer:2015knc, Das:2016vkr, Crivellin:2017zlb}. 
 The optimization of the search strategies for LQs at the LHC is thus very important to unveil the physics behind the flavor anomalies and in general for seeking BSM physics.
 
 The general LQ phenomenology at hadron colliders has been 
 explored in \cite{Mohapatra:1984aq} and more recently in \cite{Dorsner:2016wpm, Dorsner:2018ynv, Diaz:2017lit}. The relevant production mechanisms at the large hadron collider (LHC) are pair production driven by QCD interactions and single production mediated by model-dependent couplings of the LQs to leptons and quarks. An other possibility to detect the LQs is by analyzing high-$p_T$ di-lepton distributions, whose tails can be affected by the $t$-channel exchange of LQs, mediated by model-dependent couplings \cite{Greljo:2017vvb}. 
Several searches, which give bounds on the LQ masses, have been performed by ATLAS and CMS in the pair production channel at the 13 TeV LHC \cite{CMS:2018itt, Sirunyan:2018kzh, Sirunyan:2017yrk, Sirunyan:2018nkj, Aaboud:2016qeg}, while the study in \cite{CMS:2018hjx} considers the single production of third-generation LQs decaying into $b \tau$. Limits can also be obtained by recasting the results of the searches for supersymmetry. The strongest limits on 2/3-charged third-generation LQs are currently set by the CMS analysis in \cite{Sirunyan:2018kzh}, which used 35.9 fb$^{-1}$ of data at a center of mass energy $\sqrt{s}=13$ TeV.  Ref. \cite{Sirunyan:2018kzh} reinterprets the results of a search for gluinos and squarks to constraint pair produced LQs decaying into a neutrino plus a top, a bottom or a light jet. A vector LQ decaying 50\% to $t\nu$ is excluded by this analysis for masses below 1530 GeV, in the Yang-Mills (YM) case, and for masses below 1115 GeV in the minimal coupling (MC) scenario. A scalar LQ decaying 100\% to $t\nu$ is excluded up to masses of 1020 GeV. In our study we will try to improve the search strategy applied in this search and we will estimate the sensitivity of the LHC at a collision energy of 14 TeV and at high luminosity. 

 Projections of the reach of the High-Luminosity LHC (HL-LHC) and future colliders \cite{Golling:2016gvc} on different types of LQs have been presented in \cite{Allanach:2017bta}, considering pair production in the $\mu\mu jj$ channel, in \cite{Hiller:2018wbv} for pair and single production in the $b\mu\mu$ and $bb\mu\mu$ channel, and in \cite{Dumont:2016xpj} for a scalar LQ in the $bb\nu\nu$ and $cc\tau\tau$ channel. Estimates of the HL-LHC reach on LQs, based on an extrapolation of the results of current experimental searches, have been also shown in \cite{Buttazzo:2017ixm, Marzocca:2018wcf}.  A recent study has also analyzed the HL-LHC reach on a vector LQ in the $t\bar{t}$ plus missing energy channel \cite{Biswas:2018snp}. Despite considering the same final state, our analysis and search strategy will be different, relying more on the identification of the tops.  
 
In our study we consider pair produced vector and scalar LQs each decaying into a top and a neutrino, leading to a final state of two tops plus missing energy. This channel, due to a peculiar topology and to the possibility of exploiting the top tagging to disentangle the signal from the background, proves to be very powerful and, as we will show, it represents one of the best channels to probe LQs involved in the explanation of the flavor anomalies.\\
We outline a search strategy which relies in tagging the two tops in the final state, indicate the HL-LHC reach and point out several observables, that in case of a future observation of the LQ signal, can distinguish between a scalar and a vector LQ. Finally, we present implications of our results to models that explain the recent flavor anomalies. \\

The paper is organized as follows: we define our model setup in Section \ref{sec:setup}, we define our search strategy in Section \ref{sec:search} and present our results in Section \ref{sec:reach}. Shape observables to distinguish between scalar and vector LQs are shown in Section \ref{sec:scalarVSvector}. In Section \ref{sec:implications} we discuss the implications of our findings to the flavor anomalies. We offer our conclusions in Section \ref{sec:conclusions}.

\section{Setup}
\label{sec:setup}

LQ states can be classified ~\cite{Buchmuller:1986zs,Dorsner:2016wpm} in terms of their spin (scalar or vector) and SM quantum numbers, $(SU(3)_c,SU(2)_L,U(1)_Y)$, where the electric charge, $Q=Y+T_3$, is the sum of the hypercharge $(Y)$ and the third component of the weak isospin ($T_3$). In scenarios with baryon number violating couplings, these particles need to be very heavy in order to avoid the stringent limits on the proton lifetime. On the other hand, if baryon number symmetry is respected, LQ masses and couplings need to satisfy much weaker constraints, allowing them to be considerably lighter. The phenomenology of LQs with $\mathcal{O}(1~\mathrm{TeV})$ masses is very rich, including potential signatures in flavor physics observables and in the direct searches performed at the LHC.
   
\begin{figure}
\centering
\includegraphics[scale=0.8]{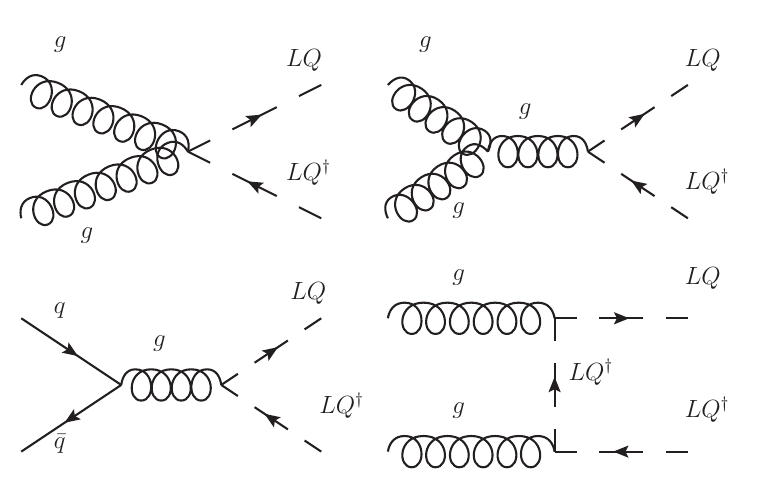} 
\caption{ \small Leading diagrams for the pair production of LQs at the LHC.}
\label{fig:diagram}
\end{figure}

In this paper we are interested in the $t\bar{t}$ plus missing energy signature at the LHC. This process can be induced by pair produced LQs, which then decay to a top quark and a neutrino. The production mechanism is dominated by gluon fusion and $q\bar q$ annihilation as illustrated in Fig.~\ref{fig:diagram}, which, for the scalar LQ, depends on a single parameter, the LQ mass. The vector LQ QCD production is controlled by a second parameter, $k$, which describes non-minimal interactions of $U_1$ with gluons and depends on underlying dynamics. The branching fraction ($\mathcal{B}$) for $\mathrm{LQ}\to t \bar{\nu}$ 
are model dependent. In Table \ref{tab:lq-states}, we list the LQ states that can decay to $t\bar{\nu}$,  along with the corresponding operator, which can arise via interactions with a lepton doublet ($L$), or a right-handed neutrino ($\nu_R$). Depending on the type of interaction and the $SU(2)_L\times U(1)_Y$ LQ representation, one can derive the maximal value of $\mathcal{B}(\mathrm{LQ}\to t\bar{\nu})$ allowed by gauge symmetry, as listed in the third column of Table~\ref{tab:lq-states}. This branching fraction can be as large as $50\%$ or $100\%$ for interactions with left-handed doublets, depending if the Yukawa coupling contributing to this also enters in the $SU(2)$ counterpart, but it can be as large as $100\%$ if interactions with right-handed neutrinos are allowed. 

Our analysis will be performed with two representative models, which can produce the same final state. Motivated by the $B$-physics anomalies, we consider: (i) the scalar LQ $S_3=(\mathbf{\bar{3}},\mathbf{3},1/3)$, and the (ii) vector LQ $U_1=(\mathbf{3},\mathbf{1},2/3)$, which we describe now in detail:

\begin{itemize}
\item[•] $\underline{S_3=(\mathbf{\bar{3}},\mathbf{3},1/3)}$: 

This $S_3$ LQ has been considered in models addressing the $B$-physics anomalies with two scalar LQs~\cite{Becirevic:2018afm,Marzocca:2018wcf}. The Yukawa Lagrangian of this model reads~\cite{Dorsner:2016wpm}
\begin{equation}
\label{eq:S3model}
\mathcal{L}_{S_3}  = y_L^{ij} \, \overline{Q^C_{i}} i \tau_2 ( \tau_k S^k_3) L_{j}+\mathrm{h.c.}\,,
\end{equation}
\noindent where $\tau_k$ ($k=1,2,3$) denote the Pauli matrices, $S_3^k$ are the LQ triplet component and $y_L$ is a generic Yukawa matrix. Note that we have neglected LQ couplings to diquarks in the above equation since they would disturb the proton stability~\cite{Dorsner:2016wpm}. An appropriate symmetry must be imposed to forbid these couplings, which are tightly constrained by experimental limits on the proton lifetime. It is convenient to recast the above expression in terms of charge eigenstates, namely,
\begin{align}
\begin{split}\label{eq:lag-S3}
\mathcal{L}_{S_3} = &- y_L^{ij} \, \overline{d^C_{L\,i}} \nu_{L\,j}\, S_3^{(1/3)}-\sqrt{2} \, y_L^{ij} \, \overline{d^C_{L\,i}} \ell_{L\,j}\, S_3^{(4/3)}\\[0.4em]
&+\sqrt{2}\,\left(V^\ast y_L\right)^{ij}\, \overline{u^C_{L\,i}} \nu_{L\,j}\, S_3^{(-2/3)}-\left(V^\ast y_L\right)^{ij} \overline{u^C_{L\,i}} \ell_{L\,j}\, S_3^{(1/3)}+\mathrm{h.c.}\,,
\end{split}
\end{align}
where $V$ is the CKM matrix. \footnote{The PMNS matrix is not relevant to our study and has been set to the identity \cite{Dorsner:2018ynv}.} The superscript denotes the electric charge of the LQ states. In this particular model, the branching fraction we are interested in reads
\begin{equation}
\label{eq:br-s3}
\mathcal{B}(S_3^{(2/3)}\to t \bar{\nu}) \simeq \dfrac{(y_L \cdot y_L^\dagger)_{33}}{\displaystyle\sum_{i}\big{(}y_L \cdot y_L^\dagger \big{)}_{ii}}\,,
\end{equation}
\noindent where we have neglected the fermion masses and used the fact that $|V_{tb}|\gg |V_{ts}| \gg |V_{td}|$. We adopted a compact notation where $(y_L \cdot y_L^\dagger)_{ii}\equiv \sum_j |y_{L}^{ij}|^2$. 

\item[•] $\underline{U_1=(\mathbf{3},\mathbf{1},2/3)}$: 

The $U_1$ model attracted a lot of attention because it can provide a simultaneous explanation to the anomalies in $b\to s$ and $b\to c$ transitions, with a single mediator~\cite{Buttazzo:2017ixm}. The most general Lagrangian consistent with the SM gauge symmetry allows couplings to both left-handed and right-handed fermions, namely,
\begin{equation}
\label{eq:lag-U1}
\mathcal{L}_{U_1} = x_L^{ij} \, \bar{Q}_i \gamma_\mu U_1^\mu L_j + x_R^{ij} \, \bar{d}_{R\,i} \gamma_\mu  U_1^\mu \ell_{R\,j}+w_R^{ij} \bar{u}_{R\,i}\gamma_\mu U_1^\mu \nu_{R\,j}+\mathrm{h.c.},
\end{equation}
\noindent where $x_L^{ij}$, $x_R^{ij}$ and $w_R^{ij}$ are Yukawa couplings. If we neglect the interactions to right-handed fields, we have, in the mass eigenstate basis:

\begin{equation}
\mathcal{L}^{L}_{U_1}= \left(V^\ast x_L\right)^{ij}  \bar{u}_{L\,i} \gamma_\mu U_1^\mu \nu_{L\,j} + x^{ij}_L \bar{d}_{L\,i} \gamma_\mu U_1^\mu \ell_{L\,j} + \mathrm{h.c.}
\end{equation}

and we obtain that
\begin{equation}
\mathcal{B}(U_1^{(2/3)}\to t \bar{\nu}) \simeq \mathcal{B}(U_1^{(2/3)}\to b \bar{\tau}) \simeq \frac{1}{2}\, \dfrac{(x_L \cdot x_L^\dagger)_{33}}{\displaystyle\sum_{i}\big{(}x_L \cdot x_L^\dagger \big{)}_{ii}}\,,
\end{equation}
where we neglected fermion masses, similarly to Eq.~\eqref{eq:br-s3}.

The $U_1$ QCD interactions that control the $U_1$ pair production are determined by the kinetic terms:
\begin{equation}
\mathcal{L}^{kin}=-\frac{1}{2} U_1^{\dagger\mu\nu} U^{1}_{\mu\nu}- i \, g_s \, k \, U^{\dagger\mu}_{1} T^{a} U_{1}^{\nu} G^a_{\mu\nu} \, ,
\end{equation}
where $U^{\mu\nu}_1$ denotes the $U_1$ strength tensor and $k$ is a dimensionless parameter which depends on the ultraviolet completion of the model. We can identify the two scenarios of minimal coupling (MC), $k=0$, and the Yang-Mills (YM) case $k=1$. 
\end{itemize} 

In the following, we will assume that the dominant interactions are the ones to third-generation left-handed fermions, as suggested by the $B$-physics anomalies. In this case, the branching fractions to $t\nu$ will be $100\%$ for $S_3$ and $50\%$ for $U_1$, which are the most optimistic values. Nonetheless, it is clear that our results can be rescaled and applied to more general flavor structures and to the other models listed in Table~\ref{tab:lq-states}.

\
 
\begin{table}[htbp!]
\renewcommand{\arraystretch}{1.7}
\centering
\begin{tabular}{|c|c|c|c|c|}\hline
Field  & Spin & Quantum Numbers &   Operators & $\mathcal{B}(\mathrm{LQ}\to t\bar{\nu})$ \\ \hline\hline
$R_2$  & $0$ & $(\mathbf{3},\mathbf{2},7/6)$ &  $\overline{u_R} R_2 i \tau_2 L$ & $\leq 0.5$ \\ 
$\widetilde{R_2}$  & $0$ & $(\mathbf{3},\mathbf{2},1/6)$ &  $\overline{Q} \widetilde{R_2} \nu_R$ & $\leq 1$ \\ 
$\bar{S}_1$  & $0$ & $(\overline{\mathbf{3}},\mathbf{1},-2/3)$ &  $\overline{u_R^C} \bar{S}_1 \nu_R$ & $\leq 1$ \\ 
$S_3$  & $0$ & $(\overline{\mathbf{3}},\mathbf{3},1/3)$ &  $\overline{Q^C} i \tau_2 \vec{\tau}\cdot \vec{S}_3L$ & $\leq 1$\\ \hline
$U_1$  & $1$ & $(\mathbf{3},\mathbf{1},2/3)$ &  $\overline{Q}\gamma_\mu U_1^\mu L\,,$~$\overline{u_R}\gamma_\mu U_1^\mu \nu_R$ & $\leq 0.5\,,~1$ \\ 
$\widetilde{V_2}$  & $1$ & $(\overline{\mathbf{3}},\mathbf{2},-1/6)$ & $\overline{u_R^C}\gamma_\mu \widetilde{V}^{\mu}_2 i \tau_2 L$\,,~$\overline{Q^C}\gamma_\mu i \tau_2 \widetilde{V}^\mu_2 \nu_R$  & $\leq 0.5\,,~1$ \\ 
$U_3$  & $1$ & $(\mathbf{3},\mathbf{3},2/3)$ & $\overline{Q}\gamma_\mu \vec{\tau}\cdot \vec{U}^\mu_3 L$  & $\leq 0.5$  \\ \hline
\end{tabular}
\caption{ \small Classification of the LQ states that can decay to $t\bar{\nu}$, in terms of the SM quantum numbers, $(SU(3)_c,SU(2)_L,Y)$, with $Q=Y+T_3$. We adopt the same notation of Ref.~\cite{Dorsner:2016wpm} and we omit color, weak isospin and flavor indices for simplicity. The last column correspond to the maximal value of $\mathcal{B}(\mathrm{LQ}\to t \bar{\nu})$, as allowed by gauge symmetries. In the cases where interactions to lepton doublets ($L$) and right-handed neutrinos ($\nu_R$) are both allowed, i.e.~for the models $U_1$ and $\widetilde{V}_2$, we give the maximal branching fraction assuming only interactions to $L$ or $\nu_R$, respectively.}
\label{tab:lq-states} 
\end{table}

\begin{figure}
\centering
\includegraphics[scale=0.6]{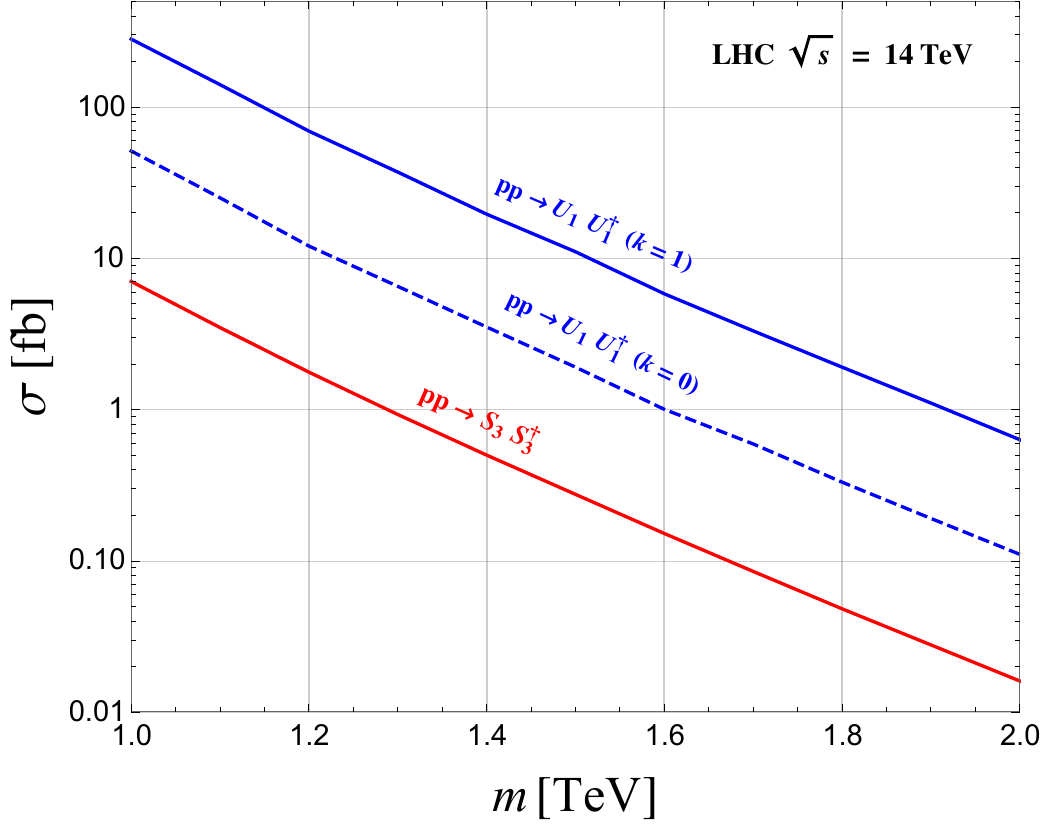} 
\caption{\small Cross section values at the 14 TeV LHC for the QCD pair production of vector (blue line) and scalar (red line) LQs. For the vector $U_1$, we indicate the cross sections in both the minimal coupling $k=0$ (dashed line) and the Yang-Mills $k=1$ (continuous line) scenarios. The cross sections are calculated at (next-to-)leading order in QCD for the (scalar) vector LQ. }
\label{fig:xsec}
\end{figure}

\section{Search strategy}
\label{sec:search}

We outline a search strategy at the 14 TeV LHC for pair-produced scalar and vector LQs, decaying each into a top quark and a neutrino. In particular, we will consider the $U_1$ and $S_3$ LQs introduced in Section \ref{sec:setup}, assuming a decay branching ratio into $t\nu$ of 50\% for $U_1$ and of $100\%$ for $S_3$. For $U_1$, we will analyze the YM scenario $k=1$. At the end, we will also present the HL-LHC reach in the MC scenario, $k=0$, which will be calculated based on the efficiencies obtained in our analysis and by rescaling the signal number of events according to the different values of the production cross sections. Fig. \ref{fig:xsec} shows cross section values at the 14 TeV LHC for the QCD pair production of $S_3$ and $U_1$ in both the YM and MC cases.

We focus on a final state given by two tops decaying hadronically plus missing energy.  
The main background consists of $Z+\mathrm{jets}$ events, where the $Z$ decays to neutrinos and leads to missing energy. Minor backgrounds, which we also include in our analysis, come from $W+\mathrm{jets}$ and $t\bar t$ events, where a leptonic decaying $W$ leads to missing energy from the neutrino and a lost lepton \cite{Sirunyan:2018kzh}. \footnote{We checked that other backgrounds, as QCD multijet events, give a negligible contribution.}

We simulate signal and background events at leading order with MadGraph5\_aMC@NLO \cite{Alwall:2014hca}. Events are then passed to Pythia \cite{Sjostrand:2006za} for showering and hadronization. We also apply a smearing to the jet momenta in order to mimic detector effects \cite{Ovyn:2009tx}. Signal events are generated via UFO files \cite{Degrande:2011ua}, created by using Feynrules \cite{Christensen:2008py}. For the case of the scalar LQ $S_3$, we apply correction factors to the cross section values, which account for QCD next-to-leading-order effects. \footnote{K-factors are included only for $S_3$. The calculations for $U_1$ are at tree-level (due to the ambiguity in the ultraviolet completion of the model).} We calculate them by using the code in Ref. \cite{Dorsner:2018ynv} (see also Ref. \cite{Mandal:2015lca}), with the choice of the NN23NLO PDF set \cite{Ball:2013hta}, $\alpha_S=0.122$ and dynamical factorization and renormalization scales. Jets are clustered with Fastjet \cite{Cacciari:2011ma} by using an anti-kt algorithm \cite{Cacciari:2008gp}. We choose a large cone size, $R=1.0$, which we identify as an optimal choice, based on the top reconstruction procedure which we will apply. \footnote{In our simulations we do not include initial state radiation or underline events. This is because, as proved for example in \cite{Sirunyan:2017nvi}, the effects of contamination on the jet invariant mass coming from these events can be eliminated by applying techniques as ``grooming" \cite{Marzani:2017mva}. We thus expect that our simulations can correctly reproduce the distributions that an experimental analysis can find after the application of these advanced jet ``cleaning" techniques.}

 The signal we want to detect is characterized by large missing transverse energy, $\slashed{E}_T$, and at least two fat-jets, coming from the hadronic decays of the two tops. Considering these signal features, as a first step of our analysis, we accept the events if they satisfy the conditions:
\begin{equation}
\slashed{E}_T > 250 \, \text{GeV}\,, \qquad n_j\geq2 \; \, (p_T \, j > 30 \,  \text{GeV} \, , \,  |\eta_j|<5)  \, ,\qquad \ \textsf{lep veto}\,,
\end{equation}
with $n_j$ denoting the number of jets satisfying the $p_T$ and rapidity requirements in the parenthesis. Events are rejected if at least one isolated lepton, either a muon or an electron, with $p_T>$ 10 GeV and in the central region $|\eta|<$ 2.5 is found (\textsf{lep veto}). \footnote{We consider the lepton isolated if it is separated from a jet by $\Delta R>$0.4. The choice of 10 GeV as a trigger on the lepton $p_T$ is a conservative choice for the evaluation of the $W$+jets and $t\bar{t}$ background contribution. Indeed, the current ATLAS trigger is 7 GeV for electrons and 6 GeV for muons \cite{Aad:2013ija}. }

\begin{figure}
\centering
\includegraphics[scale=0.4]{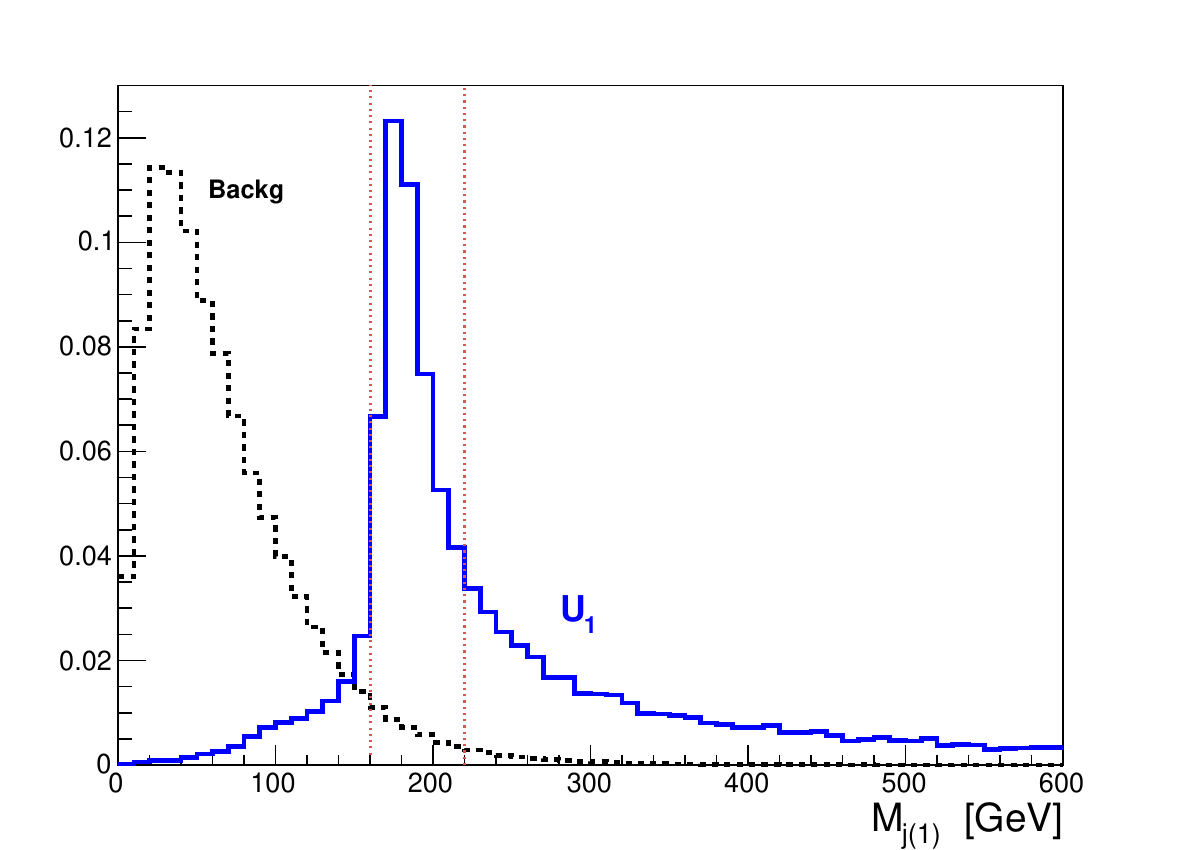}~
\includegraphics[scale=0.42]{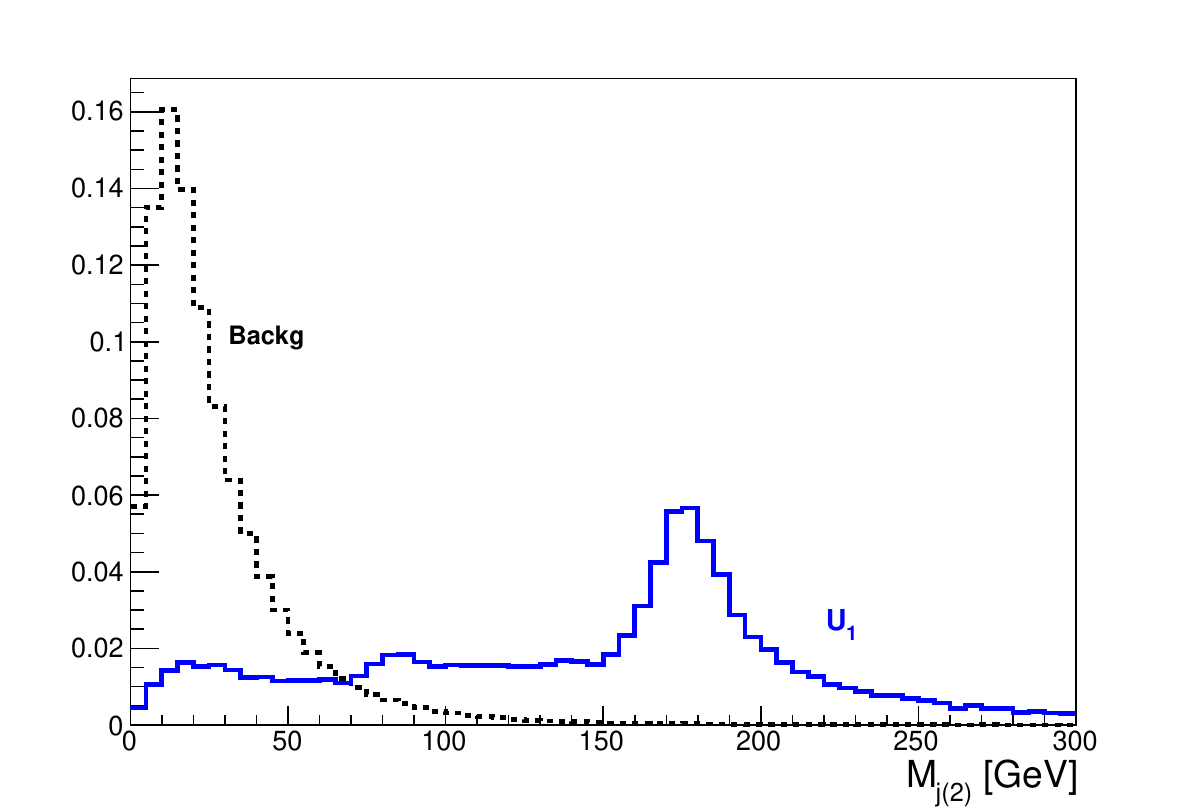}
\caption{ \small Normalized distributions for background and a signal from a LQ $U_1$ of 1.7 TeV (in the YM case). Left Plot: invariant mass of the $p_T$ leading jet. The red dotted lines mark the window around the top mass considered in this analysis.  Right plot: invariant mass of the second-leading jet.}
\label{fig:top}
\end{figure}

A crucial part of our search strategy relies on the reconstruction of both of the two tops in the final state. To reconstruct the top pair we apply the following procedure. We first consider the invariant mass of the leading jet. Since the jets are clustered on a relatively large cone size and the tops in the signal are boosted, most of the top decay products are collected in a single fat-jet. As we can see from the plot on the left in Fig. \ref{fig:top} , the invariant mass of the $p_T$-leading jet, $j(1)$, is centered around the top mass for a large portion of the signal events. As a first step of the reconstruction procedure we thus select the events with the $j(1)$ invariant mass, $M_{j(1)}$, in the region [160 GeV, 220 GeV]. $j(1)$ is then identified with the $p_T$-leading top, $t(1)$. We then analyze the invariant mass of the second-leading jet. As evident from the plot on the right in Fig. \ref{fig:top}, the majority of signal events are again centered around the top mass. If $M_{j(2)}$ is within the region [160 GeV, 220 GeV] we identify the second-leading top, $t(2)$, with $j(2)$. A small portion of signal events for $M_{j(2)}$ is centered around the $W$ mass. In order to retain these signal events we consider the $j(2)$ invariant mass region [70 GeV, 110 GeV] and the system given by the second and the third $p_T$-leading jets. If the invariant mass of the $j(2)$ plus $j(3)$ system is within the interval [160 GeV, 220 GeV] the system is identified with $t(2)$. We select the events where both of the two tops, $t(1)$ and $t(2)$, have been identified with the outlined procedure. The efficiency of our top pair tagging is of about 20\% for the signal, while we can reject the background by a factor of about 1.4$\cdot 10^3$. Table \ref{tab:acc-top-reco} indicates the cross section values for signal and background after the acceptance cuts and after the reconstruction and tagging of the pair of top quarks.

\begin{table}[h]
\begin{center}
\begin{tabular}{ | c c c c | c c c c |}
 \multicolumn{4}{c}{ Acceptance } & \multicolumn{4}{c}{ Top Tagging } \\
\hline
& &  & & & & &   \\
& &  & & & & &   \\[-0.7cm]
\multicolumn{2}{|c}{$U_1$ (YM)} & \multicolumn{2}{c|}{$S_3$} &\multicolumn{2}{c}{$U_1$ (YM)} & \multicolumn{2}{c|}{$S_3$}\\
 & &  & & & & &  \\
& &  & & & & &   \\[-0.6cm]
 $m$ [TeV] & $\sigma$ [fb] & $m$ [TeV] & $\sigma$ [fb] & $m$ [TeV] & $\sigma$ [fb] & $m$ [TeV] & $\sigma$ [fb] \\
 & &  & & & & &  \\
& &  & & & & &   \\[-0.8cm]
 1.6 & 0.45 & 1.1 & 1.4 & 1.6 & 0.097 & 1.1 & 0.23\\
 1.7 & 0.26 & 1.2 & 0.71  & 1.7 & 0.056 & 1.2 & 0.13\\
 1.8 & 0.15  & 1.3 & 0.38  & 1.8 & 0.032  & 1.3 & 0.073 \\
 1.9 & 0.084 & 1.4 &  0.21 & 1.9 & 0.019 & 1.4 & 0.042 \\
 2.0 & 0.050 & 1.5 &  0.11 & 2.0 & 0.011 & 1.5 & 0.024 \\
 2.1 &  0.030 & 1.6 &  0.064 & 2.1 & 0.0068  & 1.6 & 0.013 \\
\hline
 & &  & & & & &  \\
& &  & & & & &   \\[-0.7cm]
& & $\sigma$ [fb] & & & & $\sigma$ [fb] &\\
 & &  & & & & &  \\
& &  & & & & &   \\[-0.8cm]
& $Z+jets$ & 4560 & & & $Z+jets$ & 3.02 &\\
& $W+jets$ & 1330 & & & $W+jets$ & 0.86 &\\
& $t \bar{t}$ & 95 & & & $t \bar{t}$ & 0.36 &\\
 & &  & & & & &  \\
& &  & & & & &   \\[-0.8cm]
& Tot. Backg & 5990 & & & Tot. Backg & 4.24 & \\
 & &  & & & & & \\
\hline
\end{tabular}
\caption{\small Cross section values at the 14 TeV LHC after the acceptance cuts (columns on the left) and the identification of the pair of top quarks (columns on the right). 
\label{tab:acc-top-reco}
\small 
  }
\end{center}
\end{table}

Once having identified the tops $t (1)$ and $t(2)$,  
we construct several observables based on them, which efficiently discriminate the LQ signals from the background. 
We will thus complete our signal selection by applying cuts on these ``top observables". One of these observables, that we indicate by $M_{T2}$, is inspired by the $M_{T2}$ variable commonly used by experimental searches \cite{Sirunyan:2017kqq}. In our study it is constructed upon the tops, instead on jets, and it is defined as

\begin{align}\label{eq:MT2}
\begin{split}
& M_{T2} \equiv \text{max} \left\{ M_{T\, t(1)} , M_{T \, t(2)} \right\}\,, \\
&\\[-0.2cm]
M_{T\, t(i)} & = \sqrt{ 2   \slashed{E}_T  \,  p_T \, t(i)\,  \big( 1- \Delta\phi(\slashed{E}\, , t(i))/\pi \big) }\,, \qquad i=1,2\,,
\end{split}
\end{align}
where $p_T \, t(1,2)$ is the transverse momentum of the top $t (1,2)$ and $\Delta\phi(\slashed{E} \, , t(1,2))$ denotes the azimuthal angular separation between the missing energy vector and the top $t (1,2)$. We then consider as a signal-to-background discriminant the invariant mass of the system made of the two tops  $t(1)$ and $t(2)$, which we indicate as $M_{tt}$. 

After the top reconstruction we thus refine our signal selection by imposing the cuts:

\begin{equation}\label{eq:cut-main}
\slashed{E}_T > 500 \, \text{GeV} \quad M_{tt} > 800 \, \text{GeV} \ ,
\end{equation}

\noindent which exploits the large missing energy and the large invariant mass of the top pair system in the signal events, and the two set of cuts on the transverse momenta of the tops and on the $M_{T2}$ variable:

\begin{align}\label{eq:ref-cuts}
\begin{split}
& \text{\it loose}: \qquad M_{T2}>800 \, \text{GeV} \quad p_T \, t(1) > 500  \, \text{GeV} \quad p_T \, t(2) > 300 \,  \text{GeV} \,, \\
& \text{\it tight}: \qquad M_{T2}>1100 \, \text{GeV} \quad p_T \, t(1) > 700  \, \text{GeV} \quad p_T \, t(2) > 500 \,  \text{GeV} \,,
\end{split}
\end{align}

\noindent where the {\it loose}  ({\it tight})  selection is applied to signals with masses up to (above) 1.4 TeV.  Signal and background distributions for the relevant observables used in this analysis are shown in Fig. \ref{fig:dist-energy}.\\

Table \ref{tab:final} presents the cross section values for signal and background after the complete selection, namely after the top tagging plus the cuts in eq. (\ref{eq:cut-main}) and the {\it loose} or {\it tight} selection in (\ref{eq:ref-cuts}).

\begin{table}[]
\begin{center}
\begin{tabular}{| c c c c c c c |}
\hline
& &&&&& \\
& &&&&& \\[-0.8cm]
\multicolumn{7}{|c|}{ $U_1$ (YM)} \\
& &&&&& \\
& &&&&& \\[-1cm]
& &&&&& \\
& &&&&& \\[-0.8cm]
 $m$ [TeV]  & 1.6 & 1.7 & 1.8 & 1.9 & 2.0 & 2.1 \\
$\sigma$ [fb] & 0.047 & 0.030 & 0.019 & 0.011 & 0.0072 & 0.0045\\
\hline
& &&&&& \\
& &&&&& \\[-0.8cm]
\multicolumn{7}{|c|}{ $S_3$} \\
& &&&&& \\[-0.7cm]
&&&&&& \\
&&&&&& \\[-0.8cm]
   $m$ [TeV]  & 1.1 & 1.2 & 1.3 & 1.4 & 1.5 & 1.6 \\
$\sigma$ [fb] & 0.12 & 0.074 & 0.047 & 0.028 & 0.011 & 0.0066\\
\hline
&&&&&& \\
&&&&&& \\[-0.8cm]
Backg.   & &   \multicolumn{2}{c}{\text{\it loose}} & & \multicolumn{2}{c|}{\text{\it tight} } \\
$\sigma$ [fb]  & &   \multicolumn{2}{c}{0.25} & & \multicolumn{2}{c|}{0.080 } \\
\hline
\end{tabular}
\caption{\small Cross section values at the 14 TeV LHC after the final selection. The {\it tight} ({\it loose}) selection in eq. \eqref{eq:ref-cuts} has been applied to signals with masses above (up to) 1.4 TeV. 
\label{tab:final}
\small 
  }
\end{center}
\end{table}

\section{HL-LHC reach}
\label{sec:reach}

Based on our final results, shown in table \ref{tab:final}, we calculate the HL-LHC reach on LQs. In particular, we derive the values for the integrated luminosity needed to exclude at 95\% confidence level (CL) or to observe at 3$\sigma$ a scalar LQ $S_3$ and a vector LQ $U_1$ as a function of their mass.
The exclusion reach at 95\% CL is calculated by a goodness-of-fit test considering a Poisson distribution for the events. The 3$\sigma$ reach is estimated according to the significance level $S/\sqrt{B}$, with $S$ ($B$) the number of signal (background) events. \\

\begin{figure}
\centering
\includegraphics[width=0.5\linewidth]{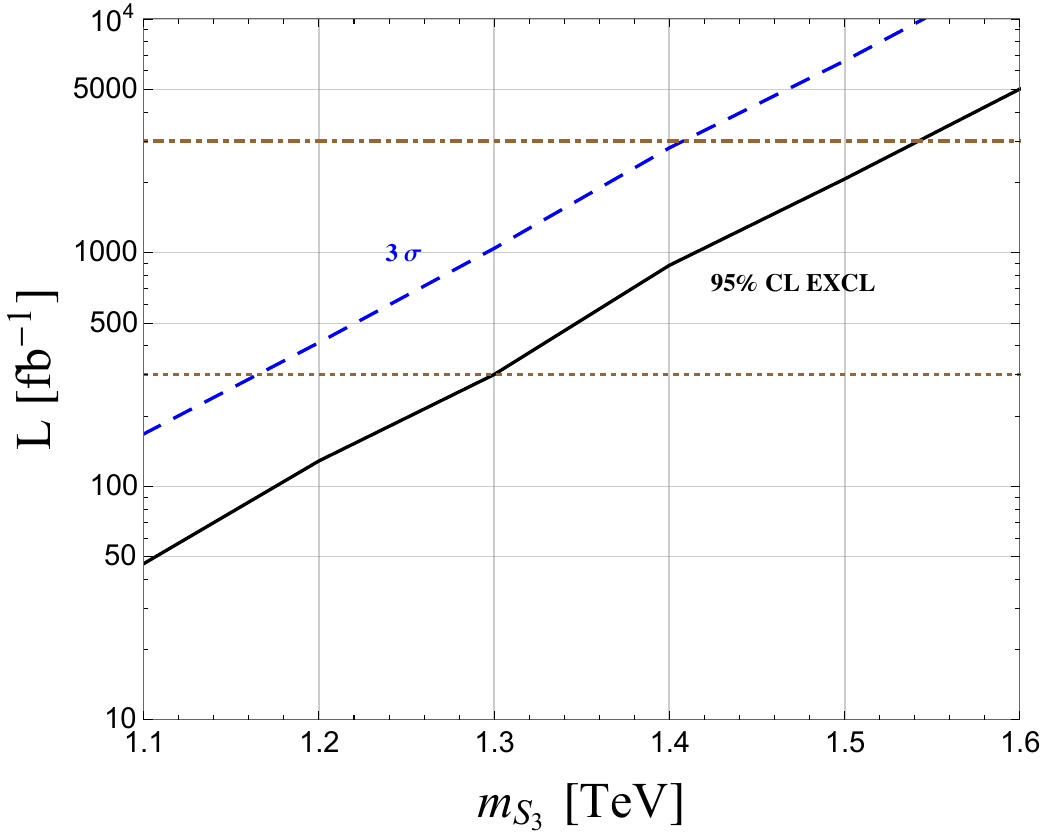} \\ [0.2cm]
\includegraphics[width=0.5\linewidth]{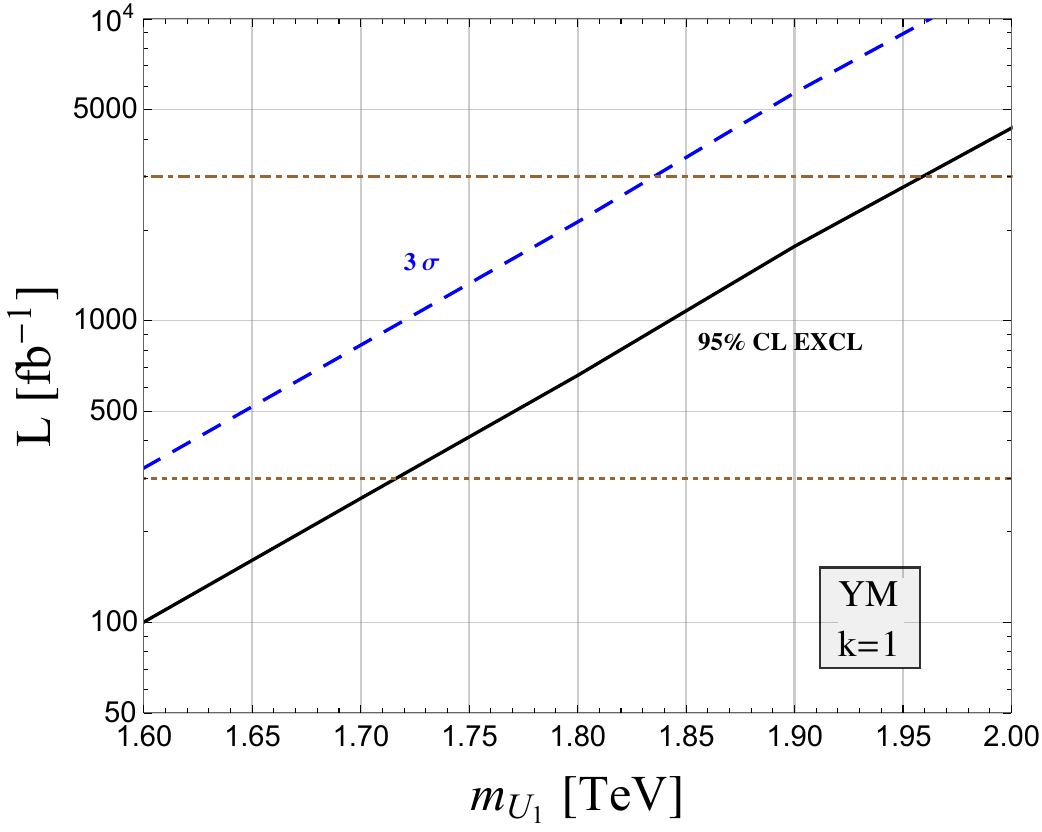}~\includegraphics[width=0.5\linewidth]{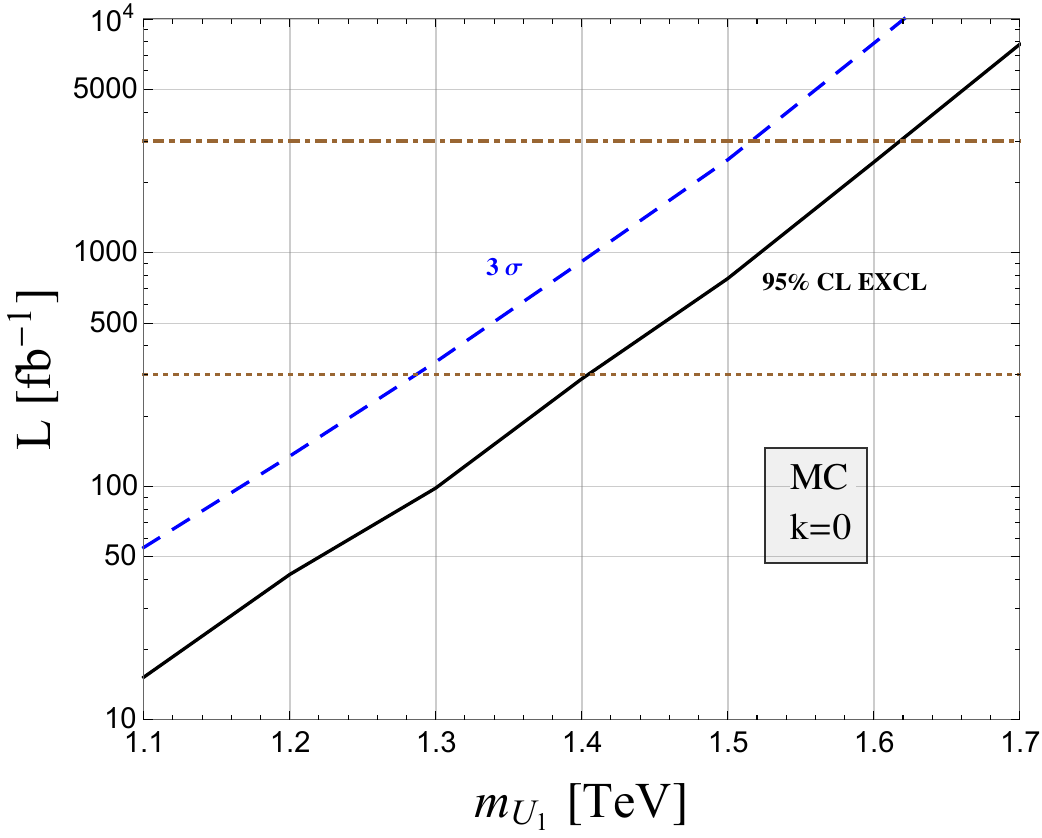} 
\caption{\small HL-LHC reach. Integrated luminosity required to exclude at 95\% C.L. (black line) or to observe at 3$\sigma$ (blue dashed line) a scalar LQ $S_3$ (upper plot) and a vector LQ $U_1$ (lower plots) as a function of their mass. For $U_1$, the plot on the left (right) refers to the YM (MC) scenario with $k=1 (0)$.}
\label{fig:reach}
\end{figure}

Fig. \ref{fig:reach} shows the HL-LHC reach on vector and scalar LQs. We see that with 3 ab$^{-1}$ (300 fb$^{-1}$) the HL-LHC can exclude a vector LQ $U_1$ up to 1.96 TeV (1.72 TeV) or observe at 3$\sigma$ the corresponding signal for masses up to 1.83 TeV (1.6 TeV) in the YM case. In the MC scenario, $U_1$ LQs  up to 1.62 TeV (1.4 TeV) can be excluded with 3 ab$^{-1}$ (300 fb$^{-1}$). For the scalar LQ $S_3$, the exclusion reach extends up to 1.54 TeV (1.3 TeV) with 3 ab$^{-1}$ (300 fb$^{-1}$), while $S_3$ as heavy as 1.41 TeV (1.16 TeV) can be observed at 3$\sigma$.

The reach for a scalar LQ $S_3$ can be confronted with the expected reach obtained by extrapolating the results of the CMS analysis \cite{Sirunyan:2018kzh}, which has been presented in \cite{Marzocca:2018wcf}. The corresponding bounds are $m_{S_3}>$ 1.43 (1.2) TeV with 3 ab$^{-1}$ (300 fb$^{-1}$). The CMS study makes use of variables as missing energy and other based on the $p_T$ of jets in the final state, but does not apply any top tagging. Given the fact that the reach of our analysis is considerably larger than the one in Ref. \cite{Marzocca:2018wcf}, \footnote{The reach in \cite{Marzocca:2018wcf} is calculated by assuming a center-of-mass energy of 13 TeV instead of 14 TeV. We calculate that the reach of our analysis remains significantly larger even if we rescale the cross sections according to the different $\sqrt{s}$.} we point out that the identification of the tops in the final state and the use of ``top observable" for the signal-to-background discrimination can improve the LHC sensitivity to LQs.
Furthermore, in our study we have applied a simple cut-and-count analysis and we expect our results to be conservative. A more refined top reconstruction, making use for example of substructure techniques as ``jettiness" \cite{Thaler:2010tr, Stewart:2010tn} or a statistical analysis of the shape of the relevant distributions considered in this study (see the subsequent Section \ref{sec:scalarVSvector}), could augment the reach of the HL-LHC. We leave these analyses to a more specialized experimental work.

\section{Distinguishing between scalar and vector LQs}
\label{sec:scalarVSvector}

We consider here the case where the HL-LHC will discover a LQ signal in the channel analyzed in this study, following the search strategy outlined in Sec. \ref{sec:search}.
 We discuss in this Section how to distinguish between the two possible signals of a vector LQ and a scalar LQ. 
 
We indicate several observables that can be used to distinguish between the two cases. A first category of observables use the difference in the energy of the final states coming from a scalar or a vector LQ. 
Indeed, due to the different scaling of the QCD pair production cross section with the mass, LQ signals from a vector LQ and a scalar one, where the vector is considerably heavier than the scalar LQ, can be identified with a similar significance.  For example, considering our results, we find that with $3$ ab$^{-1}$ a 5$\sigma$ discovery could be realized for either a vector LQ $U_1$, in the YM case, of about 1.7 TeV ($S/\sqrt{B}=5.8$) or a lighter scalar $S_3$ of about 1.3 TeV ($S/\sqrt{B}=5.1$).
The observables we identify are constructed from the reconstructed pair of tops. Tagging the two tops is thus important not only to discover the LQs but also to characterize the signal.

The ``top observables" $M_{tt}$, $M_{T2}$ and the $p_T$ of the two tops, that we already used to disentangle the signal from the background can also efficiently distinguish between $U_1$ and $S_3$.
An other observable we point out as a signal analyzer is an angular observable, specifically the azimuthal angular separation between the two reconstructed tops. $\Delta\phi (tt)$ is sensible in a non-trivial way to spin correlations \footnote{cf. \cite{Buckley:2008eb} for a study of spin correlations in the azimuthal angular separation of decay products.} and it is thus  able to probe ``directly" the spin of the LQs.\footnote{Differently from the other top observables, that can distinguish the spin ``indirectly", through the different scaling of the production cross section with the mass for vector and scalar LQs.} Similar observables, but for different topologies and constructed from the leptonic decays of tops, have been considered to identify properties of dark matter interactions \cite{Buckley:2015ctj, Haisch:2016gry} and of Higgs couplings \cite{Ellis:2013yxa, Demartin:2014fia}. 

\begin{figure}
\centering
\includegraphics[width=0.47\linewidth]{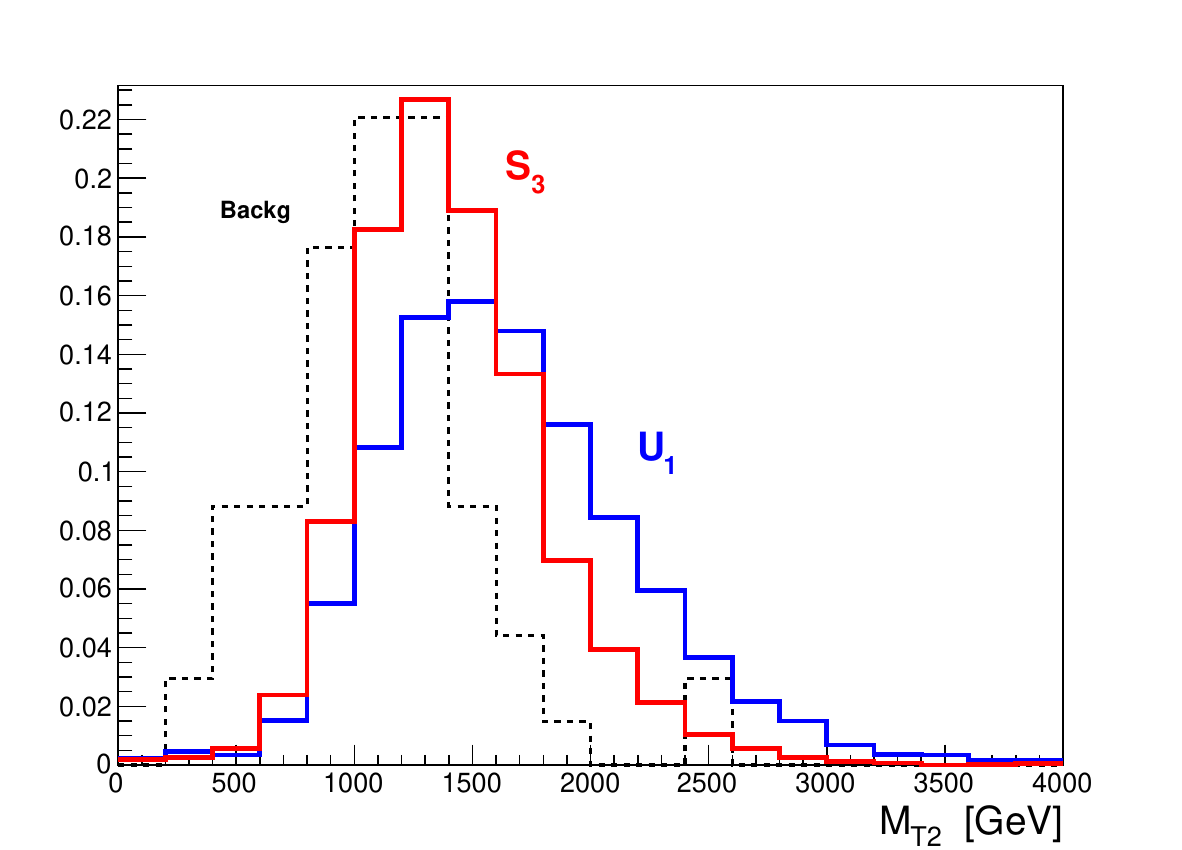}~\includegraphics[width=0.5\linewidth]{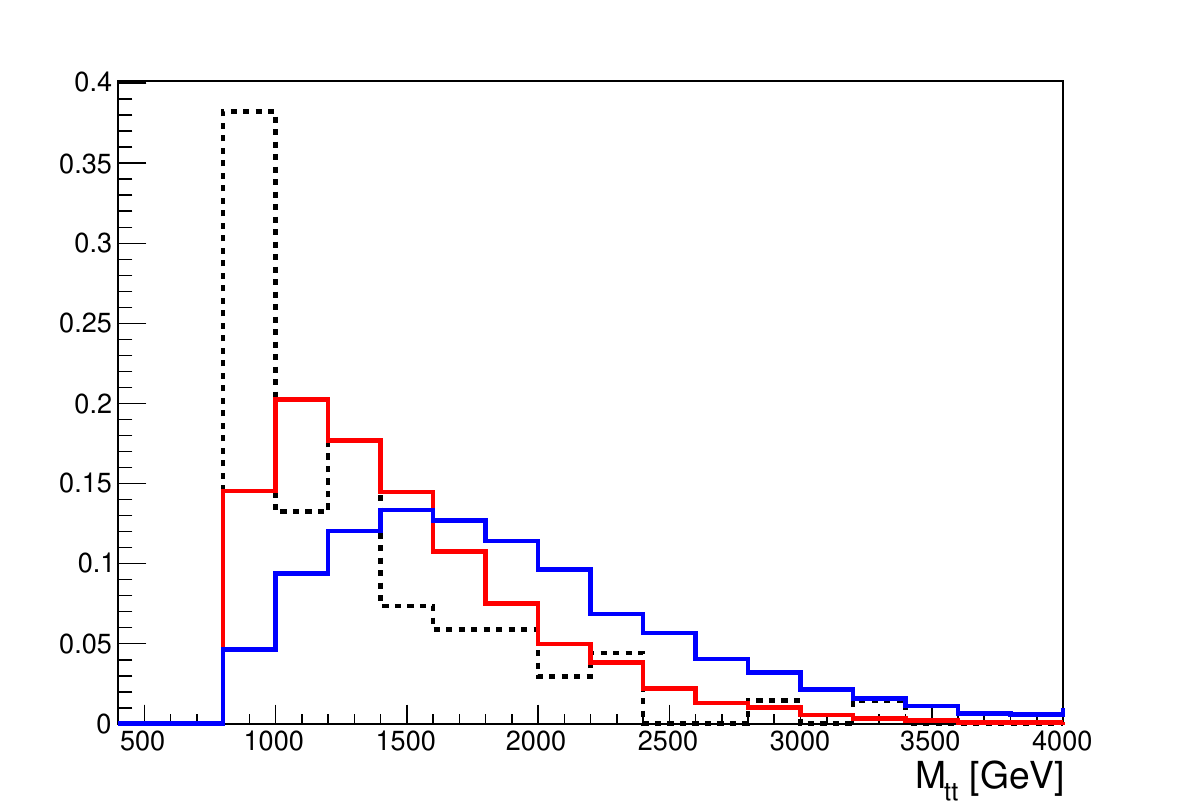}\\  
\includegraphics[width=0.5\linewidth]{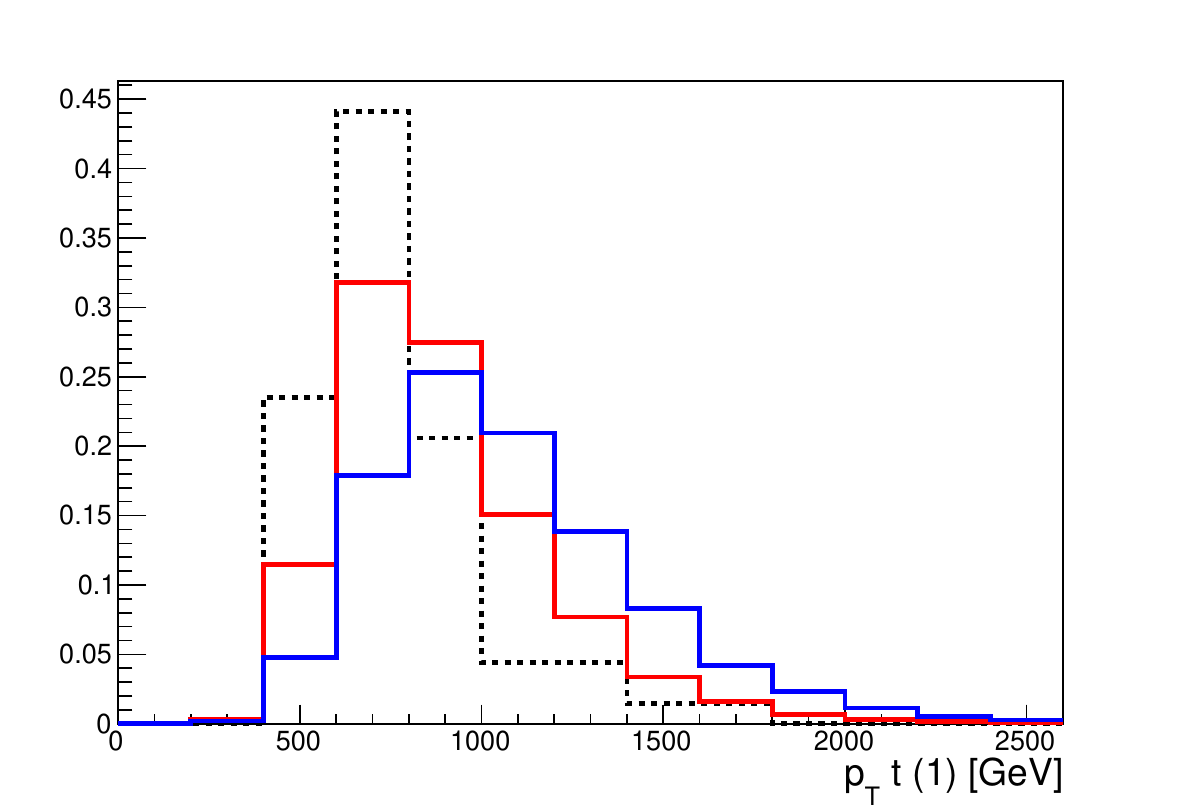}~\includegraphics[width=0.5\linewidth]{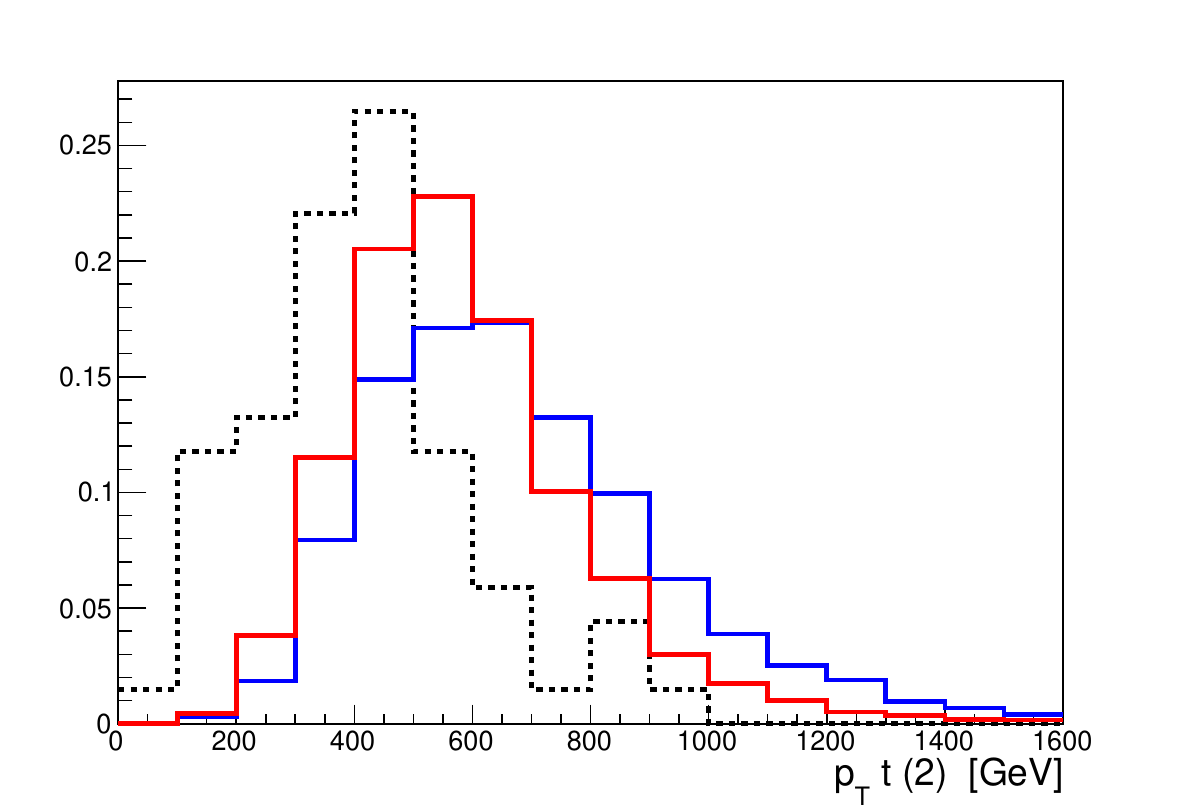}
\caption{\small Normalized distributions of ``top observables" for the background (black dotted curve) and the signals from a scalar LQ $S_3$ of 1.3 TeV (red curve) and from a vector LQ $U_1$ of 1.7 TeV with $k=1$ (blue curve). The tagging of the tops and the following cuts are applied: $\slashed{E}_T >$500 GeV, $M_{tt}>$800 GeV. Left upper plot: distributions of the $M_{T2}$ variable as defined in eq. \eqref{eq:MT2} . Right upper plot: invariant mass distributions of the reconstructed top pair. Left (right) lower plot: $p_T$ distribution of the reconstructed leading (second leading) top. }
\label{fig:dist-energy}
\end{figure}

Fig. \ref{fig:dist-energy} shows the distribution of the ``top observables" for the background and the scalar and vector LQ signals. They have been obtained after the procedure of top tagging and after applying the cuts: $\slashed{E}_T >$500 GeV, $M_{tt}>$800 GeV. We can see that the signal from $U_1$ is distributed on larger values of the top observables compared to the signal from $S_3$. The difference is particularly clear in the tails of the distributions. \footnote{In this Section we just point out the relevant observables and show the corresponding distributions. A statistical analysis based on the different shapes of the distributions is beyond the scope of the present work. }

\begin{figure}
\centering
\includegraphics[scale=0.45]{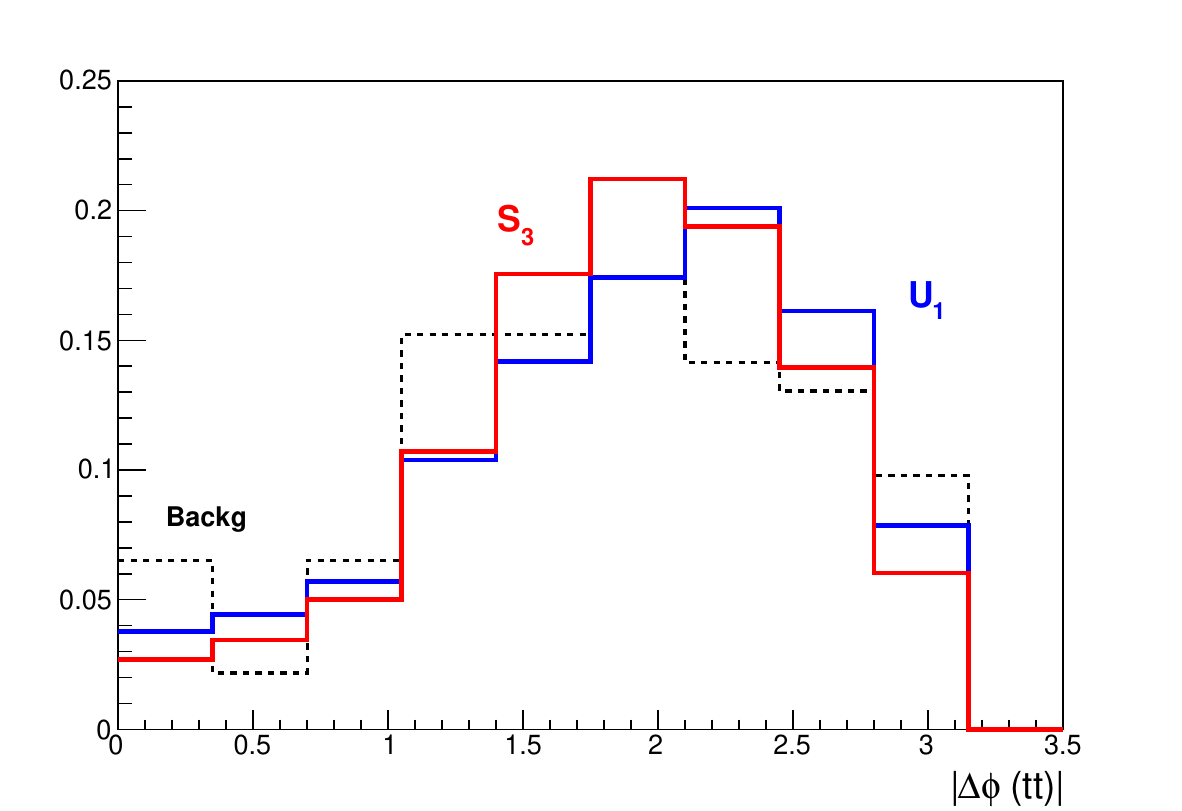}  
\caption{\small Normalized distributions of the azimuthal angular separation between the two reconstructed tops for the background (black dotted curve) and the signals from a scalar LQ $S_3$ of 1.3 TeV (red curve) and from a vector LQ $U_1$ of 1.7 TeV with $k=1$ (blue curve). The top tagging procedure and the following cuts are applied: $\slashed{E}_T >$500 GeV, $M_{tt}>$800 GeV.}
\label{fig:dist-deltaPhi}
\end{figure}

  Fig. \ref{fig:dist-deltaPhi} shows the $\Delta \phi (tt)$ distribution for the background and the $U_1$ and $S_3$ signals. The background tends to be more homogeneously distributed, the vector signal is characterized by tops at larger azimuthal separation compared to the scalar case. The difference in the $\Delta \phi (tt)$ shape depends on the spin of the LQs and does not change significantly with other values of the LQ masses, differently from the other ``top observables" discussed above. $\Delta \phi (tt)$ is thus particularly helpful to distinguish between the signals from a scalar LQ $S_3$ and a vector $U_1$ in the MC scenario $k=0$, where the differences in the energy of the final states are less marked.
  
  As a final remark, we point out that a complementary way to distinguish between different LQs could be to search for specific signatures which are present for a type of LQ and not for the others. In our scenario, for example, the vector $U_1$, differently from $S_3$, could be detectable in searches with a $b\tau$ final state, as $bb\tau\tau$, $tb\tau+ \slashed{E}_T$ or $b\tau+$ jets.

\section{Implications to the flavor anomalies}
\label{sec:implications}

LQs with masses in the TeV range are particularly interesting, since they can mediate flavor-violating processes that can accomodate the deviations from the Standard Model observed in the decays of $B$ mesons. In particular, the deviations are measured in the observables $R_{D^{(*)}}\equiv \dfrac{\mathcal{B}(B \to D^{(*)} \tau \bar\nu )}{\mathcal{B}(B \to D^{(*)} \ell \bar\nu)}$, with $\ell\equiv \mu, e$, and $R_{K^{(*)}}\equiv \dfrac{\mathcal{B}(B \to K^{(*)} \mu^+ \mu^- )}{\mathcal{B}(B \to K^{(*)} e^+ e^-)}$.

 In this Section we compare the reach on LQs derived in our study for the HL-LHC with the regions of parameters favored by the $B$-physics anomalies.

\begin{figure}[t]
\centering
\includegraphics[width=0.5\linewidth]{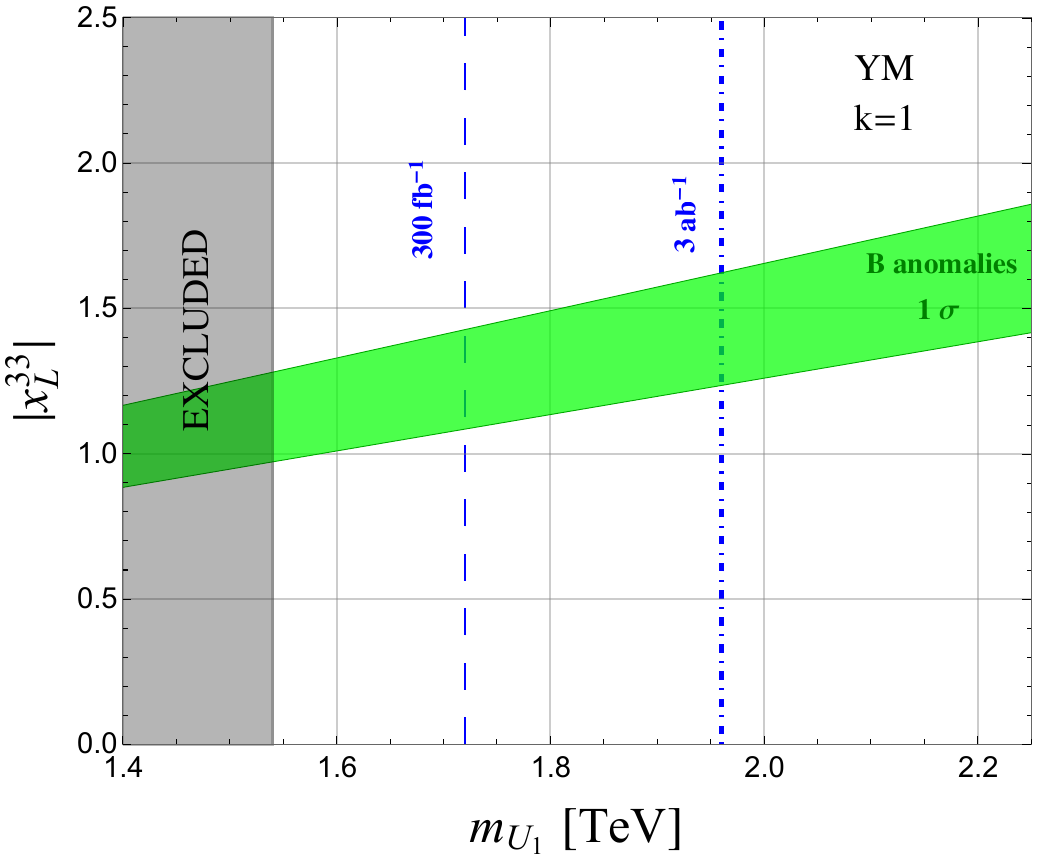}~\includegraphics[width=0.5\linewidth]{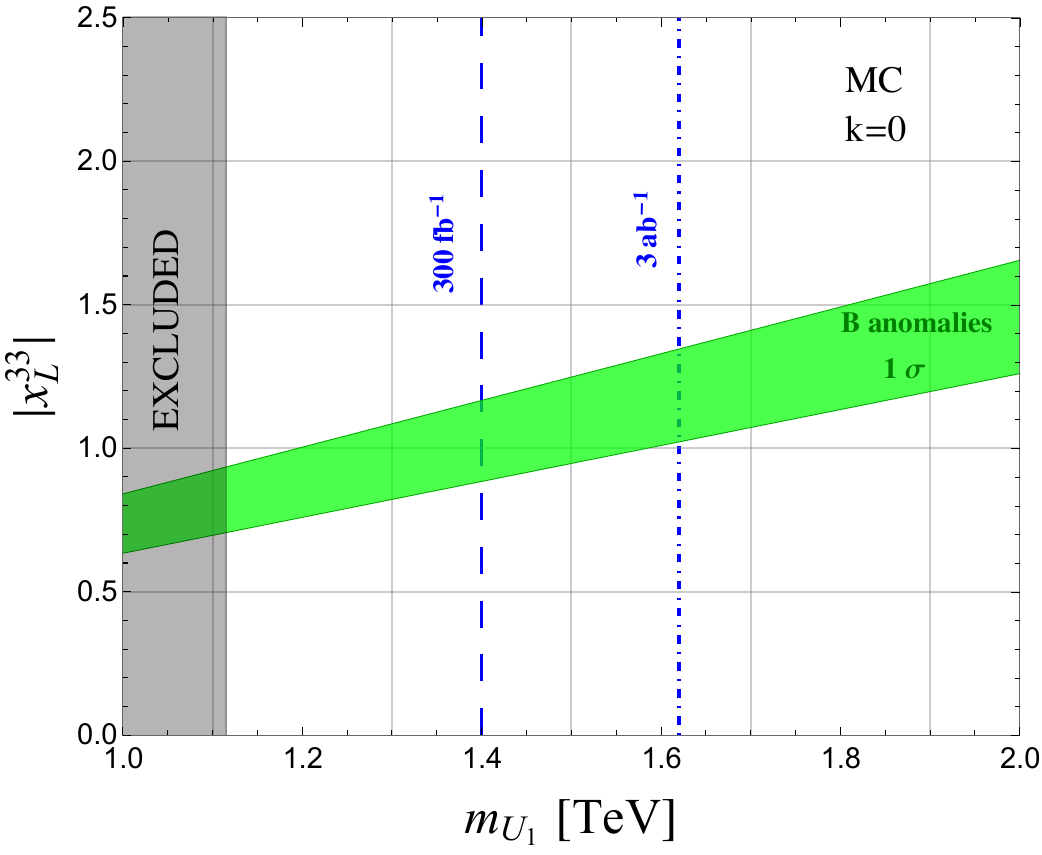} 
\caption{\small Coupling \textit{vs} mass parameter space for the vector LQ $U_1$ model preferred by $B$-physics anomalies compared to 95\% CL limits obtained in our analysis in the $t\bar{t}$ plus missing energy channel. The green band is extracted from \cite{Buttazzo:2017ixm} and represents the 1$\sigma$ region preferred by low-energy flavor observables, the gray band indicates the current limit on $m_{U_1}$, from the CMS analysis \cite{Sirunyan:2018kzh} at $\sqrt{s}=13$ TeV and 35.9 fb$^{-1}$. The blue lines are our lower bounds on the $U_1$ mass at $\sqrt{s}=14$ TeV with a luminosity of 300 (dashed line) and 3000 (dotted line) fb$^{-1}$. The plot on the left (right) refers to the YM (MC) scenario.}
\label{fig:flavor-u1}
\end{figure}

\begin{figure}
\centering
\includegraphics[width=0.5\linewidth]{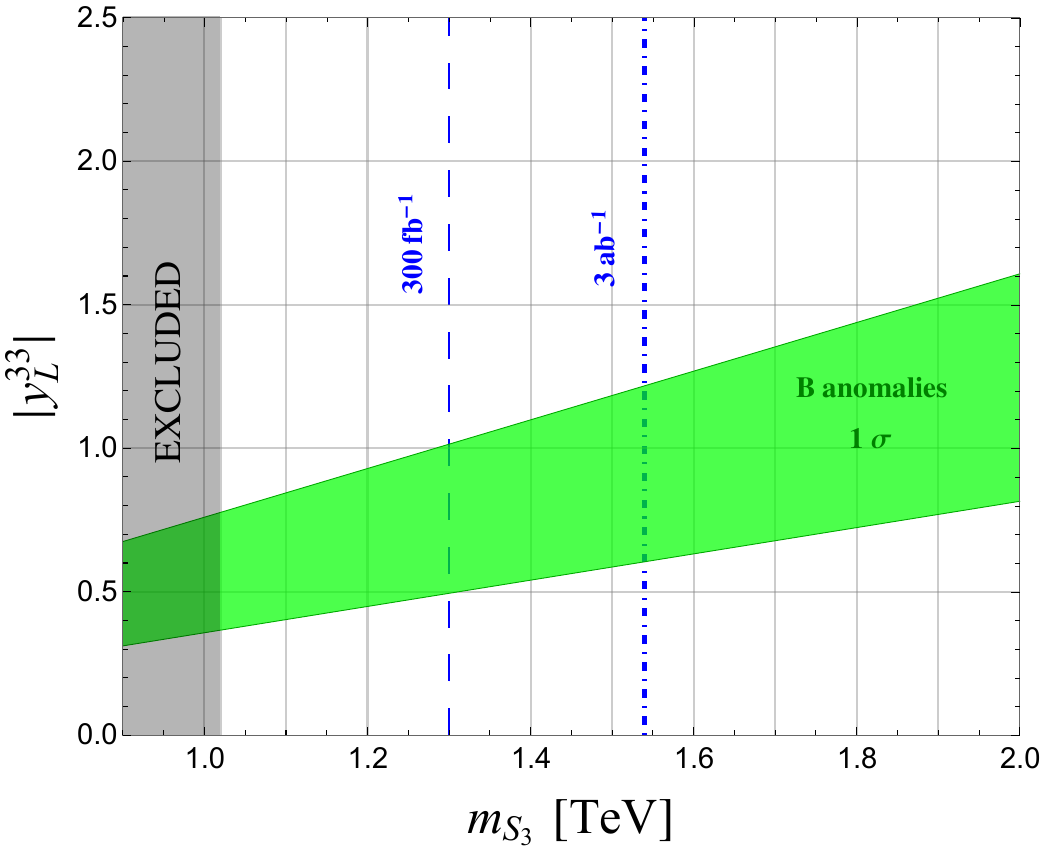}\caption{\small Coupling \textit{vs} mass parameter space for the scalar LQ $S_3$ model preferred by $B$-physics anomalies compared to 95\% CL limits obtained in our analysis in the $t\bar{t}$ plus missing energy channel. The 1$\sigma$ region preferred by low-energy flavor observables (green band) is extracted from \cite{Marzocca:2018wcf}, the gray band indicates the current limit on $m_{S_3}$, from the CMS analysis \cite{Sirunyan:2018kzh} at $\sqrt{s}=13$ TeV and 35.9 fb$^{-1}$. The blue lines are our lower bounds on the $S_3$ mass at $\sqrt{s}=14$ TeV with a luminosity of 300 (dashed line) and 3000 (dotted line) fb$^{-1}$.}
\label{fig:flavor-s3}
\end{figure}

Firstly, we consider the model with a vector $SU(2)_L$ singlet. 
This simplified model is particularly attracting because allows for a simultaneous explanation of the $B$-physics anomalies, both in $R_{D^{(*)}}$ and $R_{K^{(*)}}$, with a single particle, the LQ $U_1$ (cf.~\cite{Angelescu:2018tyl} for a recent review). 
We adopt the ansatz of Ref.~\cite{Buttazzo:2017ixm}, which is based on a $U(2)_q\times U(2)_\ell$ minimally broken flavor symmetry controlling the light generation couplings. In this case, we have, for the couplings in Eq. (\ref{eq:lag-U1}), $x_L^{ij} \equiv g_U \, \beta_{ij}$, with $\beta_{ij}=\delta_{3i}\delta_{3 j}$ up to small breaking terms of the flavor symmetry. Ref.~\cite{Buttazzo:2017ixm} calculated the region in the plane coupling \textit{vs} mass favored by $B$-physics anomalies. We show in Fig. \ref{fig:flavor-u1}  which part of this region can be probed by our analysis in the $t\bar{t}\nu\nu$ channel at the HL-LHC. The plot on the left refers to the YM scenario for the QCD pair production of $U_1$, the plot on the right to the MC case. The green band is extracted from \cite{Buttazzo:2017ixm} and represents the 1$\sigma$ region preferred by low-energy flavor observables, the gray band indicates the current limit on $m_{U_1}$, which comes from the CMS analysis \cite{Sirunyan:2018kzh} at $\sqrt{s}=13$ TeV and 35.9 fb$^{-1}$. The blue lines are our lower bounds on the $U_1$ mass at $\sqrt{s}=14$ TeV with a luminosity of 300 (dashed line) and 3000 (dotted line) fb$^{-1}$.   
We see that HL-LHC can probe in the $t\bar{t}\nu\nu$  channel a large part of the parameter space. 
Comparing with the projections presented in ~\cite{Buttazzo:2017ixm}\footnote{ Ref. \cite{Buttazzo:2017ixm} only considers the MC scenario, but the relative sensitivities of the different channels remain the same in the YM case.} for the reach in other channels, as $bb\tau\tau$, and production mechanisms, as single production, or in the di-lepton tails, we can note that our analysis in the $t\bar{t}\nu\nu$ channel appears to be the most efficient to test models with LQs involved in the explanation of the flavor anomalies.
If the HL-LHC program will exclude a LQ $U_1$ in the $t\nu$ channel, the $B$-physics anomalies can only be explained by considering large couplings $x^{33}_L>1$.\\
Finally, we consider scalar LQs. Scalar LQs, either $SU(2)_L$ triplets, $S_3$, or singlets, $S_1$, are particularly motivated  in scenarios with a BSM composite dynamics and can accomodate B anomalies \cite{Gripaios:2014tna, Marzocca:2018wcf}. In particular, $S_3$ is a good candidate to explain the anomaly in $R_{K^(*)}$ and $S_1$ the one in $R_{D^(*)}$ \cite{Angelescu:2018tyl}. 
We focus on the model in \cite{Marzocca:2018wcf}. From the $U(2)_q\times U(2)_\ell$ ansatz, the coupling in Eq. (\ref{eq:lag-S3}) reads  $y_L^{ij} \equiv g_{3} \, \beta_{ij}$, with $\beta_{ij}=\delta_{3i}\delta_{3\alpha}$. We show in Fig. \ref{fig:flavor-s3} the part of the coupling \textit{vs} $ m_{S_{3}}$ parameter space that can be probed by our results. The green band shows the 1$\sigma$ flavor fit and is extracted from Ref.~\cite{Marzocca:2018wcf}, the gray band indicates the current limit on $m_{S_{3}}$ from the CMS analysis \cite{Sirunyan:2018kzh} and the blue lines indicate the 95\% CL limits on the $S_{3}$ masses derived in our study. We see again that our analysis at the HL-LHC can probe a large portion of the parameter space relevant to B anomalies.
We can also note, by comparing our results with the projected reach in different channels shown in Fig. 2 of \cite{Marzocca:2018wcf}, that the $t\bar{t}\nu\nu$ channel is one of the most powerful to test the relevant parameter space for B anomalies.

\section{Conclusions}
\label{sec:conclusions}

We have studied the pair production of LQs in the $t \bar t$ plus missing energy channel, which is one of the most powerful to detect third-generation LQs. These particles, as highlighted in the recent literature, offer an explanation to the $B$-physics flavor anomalies. We have indicated a search strategy in the channel, which shows the advantages of tagging the tops in the final state and which uses observables constructed out of the tops to discriminate the signal from the background and to characterize the signal.  We have then assessed the reach on LQs of the future high-luminosity LHC program. Our results, presented in Fig. \ref{fig:reach}, show that with 3 ab$^{-1}$ (300 fb$^{-1}$) the HL-LHC can exclude a vector LQ, in the YM scenario, and decaying 50\% to top and neutrino, up to 1.96 TeV (1.72 TeV) or observe at 3$\sigma$ the corresponding signal for masses up to 1.83 TeV (1.6 TeV). In the MC case, a vector LQ with a mass up to 1.62 TeV (1.4 TeV) can be excluded with 3 ab$^{-1}$ (300 fb$^{-1}$). For the case of a scalar LQ completely decaying into top and neutrino, the exclusion reach extends up to 1.54 TeV (1.3 TeV) with 3 ab$^{-1}$ (300 fb$^{-1}$), while scalar LQs as heavy as 1.41 TeV (1.16 TeV) can be observed at 3$\sigma$.\\
We have further presented several observables, again constructed out of the tops, that, in case of a future discovery in the channel, can be used to distinguish between a scalar and a vector LQ (Fig.s \ref{fig:dist-energy} and \ref{fig:dist-deltaPhi}).\\
Finally, we have discussed the implications of our results to models addressing the recent  $B$-physics anomalies. The search in the $t\bar{t}\nu\nu$ channel probes to be a very efficient test of these models, with the possibility to constrain a large part of the interesting parameter space (Figs. \ref{fig:flavor-u1} and \ref{fig:flavor-s3}).

\section*{Acknowledgments}
\label{sec:acknowledgments}

This research is supported by INFN-CSN4 (HEPCUBE). The author is grateful to Olcyr Sumensari for collaboration in the early stages of the work, discussions and comments on the manuscript, and to the Mainz Institute for Theoretical Physics (MITP) for its hospitality during the completion of this work. \\

{\it Questo studio \`e dedicato alla memoria dei miei nonni Liliana, Blandina e Sante.}

\appendix


\end{document}